\documentclass[iop]{emulateapj}
 \usepackage{psfrag}
 \usepackage{rotating}
 
 \shorttitle{Mass Loss and Evolution of Classical Cepheids.}
 \shortauthors{Neilson et al.}
  
\begin{document}
  
 \title{Classical Cepheids Require Enhanced Mass Loss }
 \author{Hilding R.~Neilson\altaffilmark{1}\altaffilmark{2}}
 \altaffiltext{1}{Argelander Institute for Astronomy, University of Bonn, Auf dem H\"{u}gel 71, 53121 Bonn, Germany}
 \altaffiltext{2}{Department of Physics \& Astronomy, East Tennessee State University, Box 70652, Johnson City, TN 37614 USA}
\email{neilsonh@etsu.edu}
\author{Norbert Langer\altaffilmark{1}}

\author{Scott G.~Engle\altaffilmark{3}}
\altaffiltext{3}{Department of Astronomy \& Astrophysics, Villanova University, 800 Lancaster Ave. Villanova, PA 19085 USA}
 \author{Ed Guinan\altaffilmark{3}}
 \author{Robert Izzard \altaffilmark{1}}

% \author{et al.}
%\affil{the Land of Oz}

\begin{abstract}
Measurements of rates of period change of  Classical Cepheids probe stellar physics and evolution.  Additionally, better understanding of Cepheid structure and evolution provides greater insight into their use as standard candles and tools for measuring the Hubble constant.  Our recent study of the period change of the nearest Cepheid, Polaris, suggested that it is undergoing enhanced mass loss when compared to canonical stellar evolution model predictions.  In this work, we expand the analysis to rates of period change measured for about 200 Galactic Cepheids and compare them to population synthesis models of Cepheids including convective core overshooting and enhanced mass loss.  Rates of period change predicted from stellar evolution models without mass loss do not agree with observed rates whereas including enhanced mass loss yields predicted rates in better agreement with observations.  This is the first evidence that enhanced mass loss as suggested previously for Polaris and $\delta$ Cephei must be a ubiquitous property of Classical Cepheids.
\end{abstract}

\section{Introduction}
Observations of real-time stellar evolution are rare and difficult to obtain as most stars change on time scales much longer than a human life time.  However, stellar pulsation properties are observed to change. Just as \cite{Eddington1926} demonstrated that observations of stellar pulsation and the pulsation period constrain the interior structure of a star via the period-mean density relation, changes in the pulsation period constrain the evolution of that star as changes in the mean density.  Currently, period change has been measured for about 200 Classical Cepheids \citep{Turner2006} as well as a number of Type II Cepheids \citep{Schmidt2004, Schmidt2005, Rabidoux2010} and RR Lyrae stars \citep{LeBorgne2007, Vandenbroere2012}.

Rates of period change constrain which stage a Cepheid is evolving: a negative rate of period change suggests a Cepheid is evolving blueward on the Hertzsprung-Russell diagram on its second crossing of the Cepheid instability strip, while a positive rate suggests redward evolution on the first or third crossing.  The first crossing occurs when a star evolves through the Hertzsprung gap at the end of main-sequence evolution.  A star cools as it crosses the Hertzsprung gap on a thermal time scale hence crosses the Cepheid instability strip more quickly than in later stellar evolutionary phases which are on a nuclear timescale. The rate of period change of a Cepheid in this phase of evolution is positive and greater than rates of period change during later crossings of the instability strip. The third crossing occurs near the end of helium-core burning when the star evolves to the asymptotic giant branch or to the red supergiant branch. 

\cite{Turner2006} found that measured pulsation periods and rates of period change appear consistent with predictions from stellar evolution models.  However, \cite{Neilson2012a} computed a grid of stellar evolution models to fit the observed effective temperature and luminosity of Polaris but could not simultaneously fit its observed rate of period change.  Predicted rates of period change consistent with the observed luminosity and effective temperature are too small by more than $1~$s~yr$^{-1}$. Stellar mass loss at the rate of $\dot{M} \sim 10^{-6}~M_\odot~\rm{yr}^{-1}$ was suggested to account for the difference between theoretical and observed period changes.  This hypothesis is contentious, as the necessary mass loss rate is rather high, but if correct is an important key to understanding Cepheid structure and evolution.

Cepheid mass loss is an important ingredient for understanding late-stage stellar evolution, the transitions from red-to-blue supergiants \citep{Mackey2012} and to solve the long-standing problem of the Cepheid mass discrepancy \citep{Cox1980, Brocato2004, Caputo2005, Bono2006,Keller2008, Neilson2011}.  The Cepheid mass discrepancy is the difference between Cepheid mass predictions  using stellar evolution models and stellar pulsation models, where the latter predicts masses approximately $10$-$20\%$ smaller.  \cite{Bono2006} suggested a number of potential solutions to resolve the mass discrepancy with the two most promising ones being Cepheid mass loss and convective core overshooting in main-sequence Cepheid progenitors.  Mass loss removes mass from the stellar envelope whereas overshooting leads to a more luminous Cepheid for a given initial stellar mass.  

\cite{Keller2008} argued that the mass discrepancy increases for decreasing metallicity, which is inconsistent with radiative-driven mass loss theory, hence advocating convective core overshooting as the solution.  On the other hand, \cite{Neilson2008, Neilson2009a} developed a pulsation-driven mass loss model and found that pulsation-driven mass loss does not decrease with decreasing metallicity. \cite{Neilson2011} added the pulsation-driven mass-loss theory to the stellar evolution models and found that a combination of overshooting and Cepheid mass loss is required to account for the mass discrepancy.

Infrared and radio observations also provide evidence for Cepheid mass loss. Interferometric observations of a number of Cepheids suggest $K$-band flux excess \citep{Kervella2006, Merand2006, Merand2007}, indicating the presence of circumstellar material from a stellar wind, but no flux excess about the non-pulsating yellow supergiant $\alpha$ Persei.  Infrared excess was also found in {\it Spitzer} observations  of Galactic Cepheids \citep{Marengo2010,Barmby2011}, in particular,  observations resolve a bow shock about the Cepheid prototype $\delta$ Cephei. \cite{Neilson2009b,Neilson2010} also found evidence for infrared excess due to mass loss in Large Magellanic Cloud Cepheids.  Furthermore, \cite{Matthews2012} detected significant mass loss in $\delta$ Cephei based on 21~cm radio observations, including a measurement of the wind velocity.  All of these analyses suggest mass-loss rates of the order $10^{-8}$ - $10^{-6}~M_\odot~\rm{yr}^{-1}$, but could be explained by other causes such as material left over from other stages of evolution.  These mass-loss rates are much larger than is expected from a radiatively-driven wind or a Reimers relation \citep{Neilson2008}.

Measurements of the period change of Polaris provided a first direct test of Cepheid mass loss.  Whereas observational evidence for Cepheid mass loss is based on flux excess, the analysis of the rate of period change measures how mass loss changes the evolution of the Cepheid.  Period change also varies differently as a function of convective core overshoot and mass loss, providing the first method to constrain different solutions to the mass discrepancy \citep{Neilson2012a}.  However, measuring mass loss from period change for Polaris is not reliable because Polaris is only one star and its large rate of period change may be explained by other phenomena, such as magnetic fields \citep{Stothers2009}.  If mass loss is so significant then there must be some evidence for it in the measured period changes of all Galactic Cepheids \citep{Szabados1983, Turner2006}.

\cite{Neilson2008} and \cite{Neilson2012a} showed that the rate of period change is linearly proportional to the mass-loss rate, i.e.~if mass loss increases then so too must the rate of period change. If Cepheid mass loss is ubiquitous then one would expect stellar evolution models to underestimate rates of period change relative to the measured rates as well as to predict more Cepheids with a negative rate of period change than a positive rate.  \cite{Turner2006}  did not find evidence for the first hypothesis, however they did not conduct a quantitative comparison.  The purpose of this article is to compare the measured rates of period changes from \cite{Turner2006} to predicted rates computed from population synthesis models of Galactic Cepheids to test whether evolution models are consistent with the observed ratio of Cepheids with positive and negative period change. In Sect.~2, we describe the method for computing population synthesis models and the underlying model assumptions.  In Sect.~3 we compare the measured rates of period change with stellar evolution models assuming no significant Cepheid mass loss and repeat the comparison in Sect.~4 but for evolution models with significant Cepheid mass loss.  We discuss the implications of this work and conclude in Sect.~5.

\section{Methods}
We compute detailed stellar evolution models using the \cite{Yoon2005} code, as in our previous work \citep{Neilson2012a}.  Here, we use that grid of models with masses, $M = 3.2$ - $6.8~M_\odot$, in steps of $0.1~M_\odot$, with convective core overshooting parameter $\alpha_c = 0$ - $0.4$ in steps of $0.1$ and assuming the \cite{Grevesse1996} standard solar metallicity.  Convective core overshooting is discussed in \cite{Neilson2011, Neilson2012a}.  We extend this grid up to $14~M_\odot$. Rates of period change are computed from the stellar evolution tracks when the star's effective temperature and luminosity are consistent with crossing the Cepheid instability strip \citep{Bono2000} and for luminosities, $\log L/L_\odot \ge 3$, where rates are computed using Equation 2 from \cite{Neilson2012a},
\begin{equation}\label{pdot}
\frac{\dot{P}}{P} = -\frac{4}{7}\frac{\dot{M}}{M} + \frac{6}{7}\frac{\dot{L}}{L} - \frac{24}{7}\frac{\dot{T}_{\rm{eff}}}{T_{\rm{eff}}},
\end{equation}
where $\dot{M}$ is the change of stellar mass.

We then compute population synthesis models for Galactic Cepheids, assuming a \cite{Kroupa2001} initial mass function and a constant star formation rate.  From our models, we compute the relative probability of seeing a Cepheid with luminosity within a bin $L + dL$ and $L - dL$ and period change less than $\dot{P} + d\dot{P}$ and greater than $\dot{P} - d\dot{P}$.  By summing all the Cepheids with positive period change and those with negative period change, we compute the relative fraction of each that can be compared to observations.

This comparison allows us to test the hypothesis that Cepheids undergo enhanced mass loss.  Because mass loss increases the rate of period change then one would intuitively expect population synthesis models would predict that the fraction of Cepheids with positive period change to be less than the observed fraction from \cite{Turner2006}.  However, mass loss may also change evolutionary time scales, which would also change the relative numbers.  In the \cite{Turner2006} sample, there are 196 Galactic Cepheids with measured rates of period change, of which 128 Cepheids are measured to have positive rates of period change and 68 negative rates of period change.  Therefore, approximately 2/3 of all Cepheids have a positive rate of change.

\begin{figure}[t]
\begin{center}
\plotone{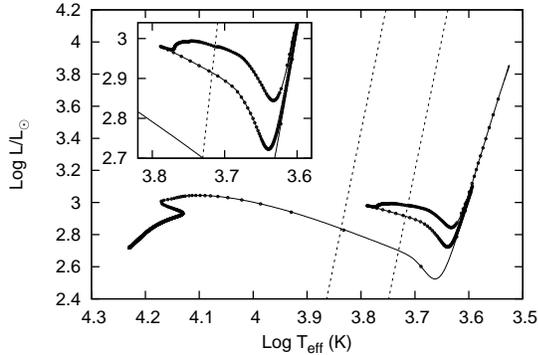}
\end{center}
\caption{The Hertzsprung-Russell diagram for a $5~M_\odot$ stellar evolution track from our sample where the points represent time steps of $10^5$~yrs of stellar evolution. Dotted lines represent the boundaries of the Cepheid instability strip \citep{Bono2000}.}
\label{f2}
\end{figure}

\section{Population Synthesis Models Without Cepheid Mass Loss}

We compute rates of period change for our grids of stellar evolution models.  The largest positive rates of period change are for stars on the first crossing during which the crossing time scale is proportional the the thermal time scale of the star.  The second and third crossings take much longer, which lead to smaller absolute values for the rate of period change. We show the time scales of evolution for a $5~M_\odot$ model in Fig.~\ref{f2}, in which we plot the Hertzsprung-Russell diagram of the track and plot points at every time step of $10^5~$yr.  The star evolves along the first crossing in about $1~$Myr and the third crossing appears to be much longer than the second crossing, in apparent contradiction to our intuitive hypothesis that there would be fewer Cepheids with positive rates of period change, i.e. third-crossing Cepheids.  

\begin{figure*}[t]
\begin{center}
\plottwo{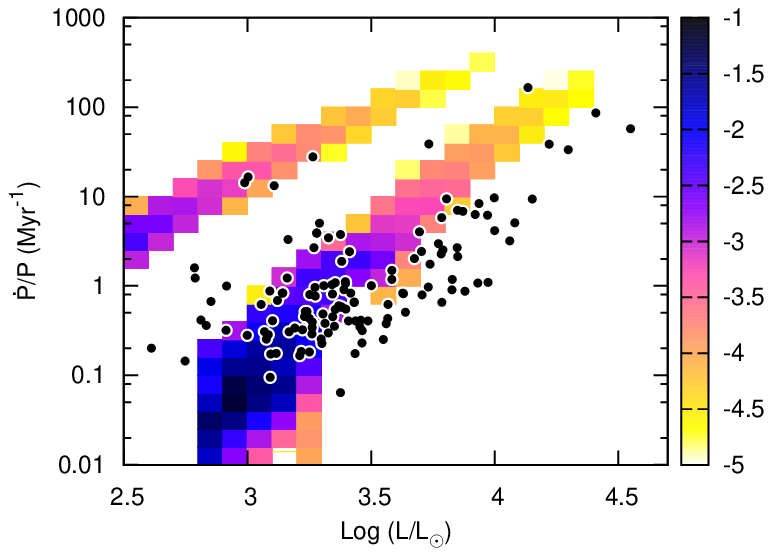}{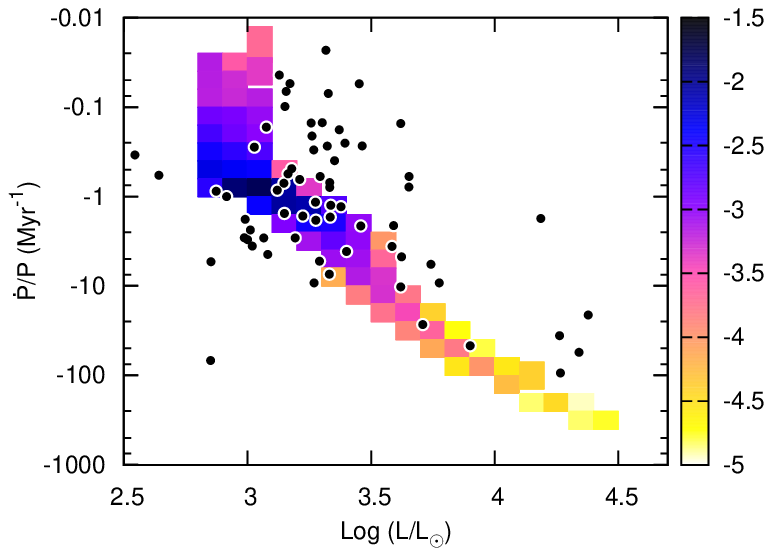}
\end{center}
\caption{The predicted positive (left) and negative (right) rates of period change as a function of stellar luminosity.  Color contours represent the logarithm of the probability of finding a Cepheid with the given rate of period change and luminosity assuming a constant star formation rate. Black dots represent rates of period change for Galactic Cepheids reported by \cite{Turner2006}.}
\label{f3}
\end{figure*}

We explore this result further by predicting the probability for observing a Cepheid with a positive or a negative rate of period change as a function of luminosity from stellar evolution models without convective core overshooting and plot the results in Fig.~\ref{f3}.  There are three apparent tracks in the two plots representing the first and third crossings of the instability strips in the left plot and the second crossing in the right plot.  Measured rates of period change for Galactic Cepheids \citep{Turner2006} are plotted for comparison.  \cite{Turner2006} presented the measurements as $\dot{P} = f( P )$, which we use in this work. However, we convert the pulsation periods to bolometric luminosities using the relation from \cite{Turner2010} for more direct comparisons with stellar evolution models as $\dot{P}/P = f(L/L_\odot)$. 

\begin{figure*}[t!]
\begin{center}
\plottwo{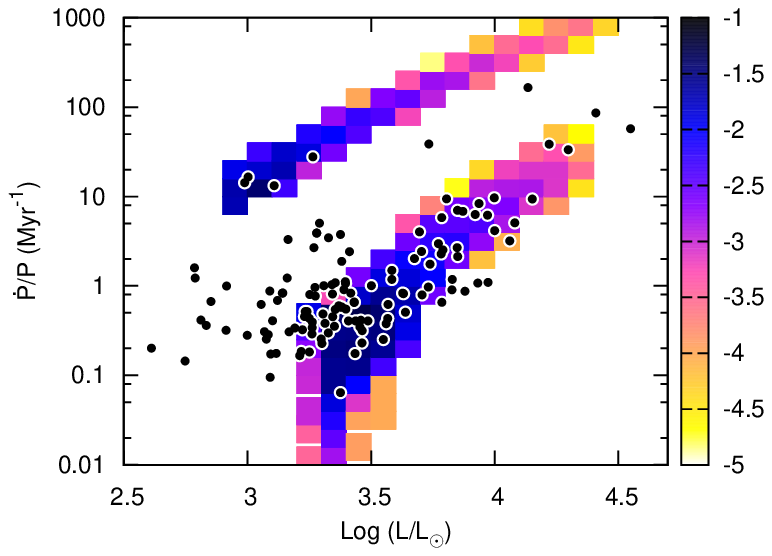}{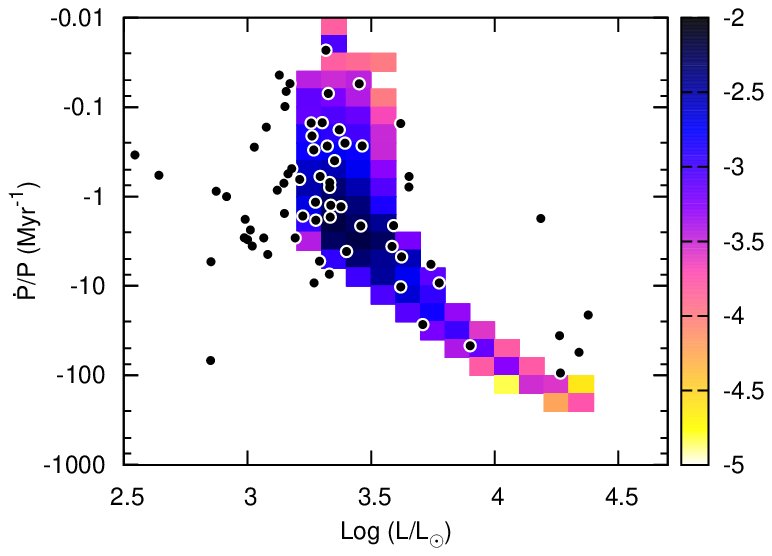}
\end{center}
\caption{Same as Fig.~\ref{f3} except using stellar evolution models with $\alpha_c = 0.2$.}
\label{f4}
\end{figure*}

The observed rates of period change do not agree well with predicted rates for a given luminosity: predicted positive rates tend to be larger than the observed rates, especially for luminosities $\log L/L_\odot > 3.5$ while observed negative rates have more scatter than predicted rates.  This difference is resolved if we employ stellar evolution models that assume a moderate amount of convective core overshoot, $\alpha_c = 0.2$.  Convective core overshoot acts to shift the contours to greater luminosities but maintaining the same behavior.  We plot the rates of period change for the $\alpha_c = 0.2$ stellar evolution models in Fig.~\ref{f4}.  The predicted positive rates of period change for models with overshoot agree better with measured rates for $\log L/L_\odot \gtrsim 3.5$, but disagree with period changes for Cepheids with smaller luminosities.  Including convective core overshoot in stellar evolution models increases a Cepheid's luminosity but the rates of period change do not appear to change.

We calculate the fraction of Cepheids that have positive period change and the fraction with negative period change from the synthesis models.  For the stellar evolution models without overshooting, we find that about $85\%$ of Cepheids have positive rates of period change.  We find a similar fraction of Cepheids from the stellar evolution models that include convective core overshoot.  These two results differ from the observed fraction of Cepheids with positive rates of period change, about $67\%$ and suggest that convective core overshoot does not affect the time scale for a Cepheid evolving redward relative to its time scale for the second crossing.  A different physical process is needed.

While it is expected that Cepheid mass loss increases the relative fraction of Cepheids with positive rates of period change, this assumes that mass loss, like convective core overshoot, does not affect the relative evolutionary time scales of the second and third crossings for a given mass.  Previous tests suggest that mass loss does, in fact, change the relative time scales as well as the shape of the blue loop on the Hertzsprung-Russell diagram \citep{Neilson2011, Matthews2012}. It is necessary to  test whether Cepheid mass loss increases or decreases the predicted fractions of Cepheids with positive and negative rates of period change by changing the relative evolutionary time scales of blueward and redward evolution.

\section{Population Synthesis Models With Enhanced Cepheid Mass Loss}
Convective core overshoot does not resolve the discrepancy between the predicted and observed ratios of the number of Cepheids with positive rates of period change and negative rates of period change.  We repeat the analysis from the previous section with a grid of stellar evolution models computed assuming no convective core overshoot while including enhanced mass loss at a rate $\dot{M} = 10^{-7}~M_\odot$~yr$^{-1}$ when the stars evolve across the Cepheid instability strip.

\begin{figure*}[t!]
\begin{center}
\plottwo{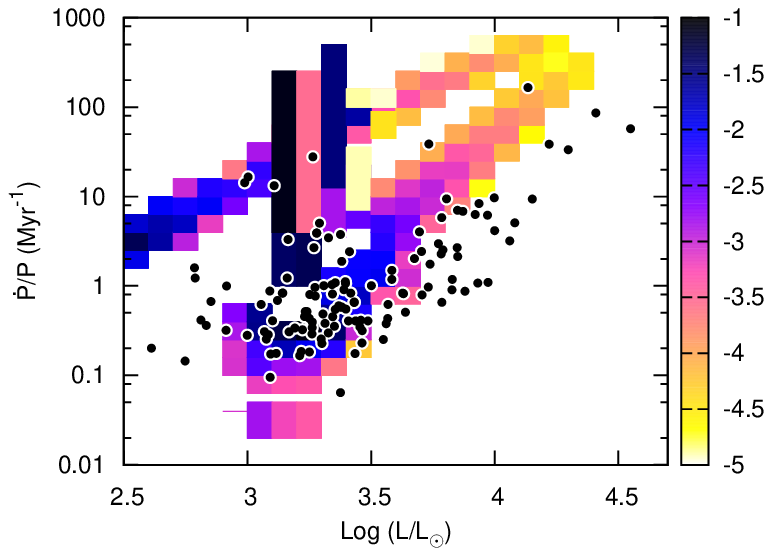}{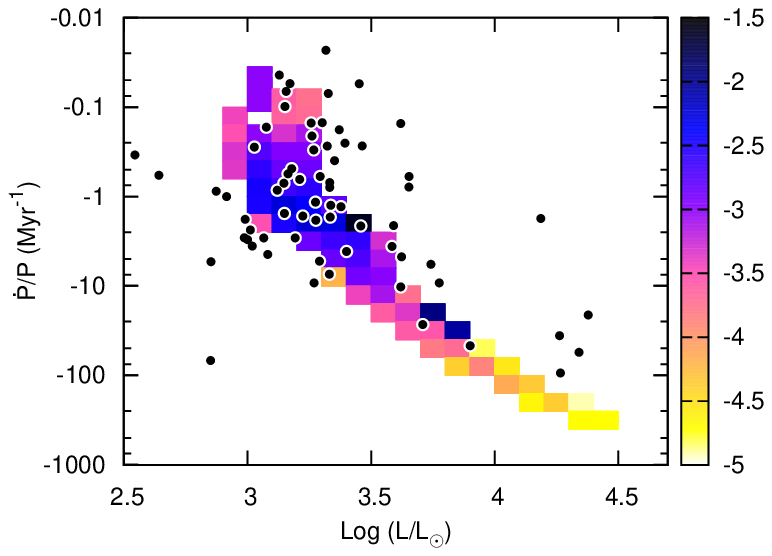}
\end{center}
\caption{Same as Fig.~\ref{f3} except using stellar evolution models with mass-loss rates $\dot{M} = 10^{-7}~M_\odot~$yr$^{-1}$ while evolving on the Cepheid instability strip.}
\label{f5}
\end{figure*}

We show in Fig.~\ref{f5} the rates of period change plotted as a function of stellar luminosity for the stellar evolution models with enhanced Cepheid mass loss.  Period change predictions for the second crossing do not change significantly (Fig.~\ref{f5}, right panel) because of the enhanced mass loss.  However, predicted positive rates of period change differ, especially rates of period change of Cepheids evolving along the third crossing of the instability strip.  Cepheid models evolving on the third crossing with luminosities $\log L/L_\odot = 3.2$ - $3.5$ have rates of period change up to one to two orders-of-magnitude greater than those predicted from models assuming no enhanced mass loss. The differences at larger luminosities may be resolved by including convective core overshooting in the stellar evolution calculations, as shown in Fig.~\ref{f4}.  As mass loss appears to increase the rate of period change then massive Cepheids could have smaller mass-loss rates.

From the models including enhanced mass loss,  71\% of the Cepheids have  positive rates of period change, a significantly smaller fraction than that predicted from stellar evolution models without mass loss enhancement, 85\%, and in better agreement with that measured from observations, 67\%.  This result suggests that theoretical stellar evolution models require enhanced mass loss on the Cepheid instability strip to predict rates of period change consistent with observations.

\section{Discussion}
We computed grids of stellar evolution models for three cases: 1) with no convective core overshooting and no enhanced Cepheid mass loss, 2) convective core overshooting, $\alpha_c = 0.2$ and no enhanced mass loss and 3) no overshooting but with enhanced Cepheid mass loss, $\dot{M} = 10^{-7}~M_\odot~$yr$^{-1}$.  For each grid, population synthesis models are used to predict the fraction of Cepheids with positive rates of period change and those with negative rates of period change.   These predicted fractions are compared to the fraction of Cepheids with positive and negative rates measured from observations \citep{Turner2006}.  

In the first two cases  our models overestimate the fraction of Cepheids with positive rates of period change and convective core overshooting does not appear to change the fraction.  However, in the third case,  the fraction of Cepheids with positive period change is smaller, improving agreement with the measured fraction. We summarize the predicted ratios in Tab.~\ref{t1}. This result is the first direct suggestion that Cepheids, in general, must undergo enhanced mass loss.  Mass loss has been inferred from excess flux in the infrared and radio observations, but these could also be interpreted as a relic from previous epochs of stellar evolution.  The velocity of the circumstellar medium  of $\delta$~Cephei  was also measured from radio observations, suggestive of  but not a direct measurement of a Cepheid wind.  
\begin{table}[t]
\caption{Fractions of Cepheids with positive and negative rates of period change}\label{t1}
\begin{center}
\begin{tabular}{lcc}
\hline
\hline
Case &Positive & Negative \\
\hline
1\footnote{Models without overshoot or enhanced mass loss.}& 85\%&15\%\\
2\footnote{Models with overshoot but no enhanced mass loss.}& 85\%&15\%\\
3\footnote{Models with enhanced mass loss but no overshoot.}&71\% & 29\%\\
Observed& 67\% & 33\% \\
\hline 
\end{tabular}
\end{center}
\end{table}

The strongest evidence for Cepheid mass loss would be the discovery of a P Cygni profile from spectral observations. P Cygni line profiles may not be detectable in Cepheids as the spectral lines are dominated by pulsation and turbulence \citep{Bersier1993}. Therefore, rates of period change appear to provide the most direct evidence for enhanced mass loss in Cepheids.

In this analysis we assumed a mass-loss rate $\dot{M} = 10^{-7}~M_\odot$~yr$^{-1}$, roughly consistent with the predictions from \cite{Neilson2008,Neilson2009a} and \cite{Neilson2011}, but without physical justification.  This particular average mass-loss rate leads to a more significant increase of the rate of period change in dimmer, i.e less massive $M \lesssim 7~M_\odot$, Cepheids but the change is negligible at higher mass. We call this the average mass-loss rate because \cite{Neilson2011} suggested that mass loss may change by orders of magnitude as a Cepheid evolves across the instability strip. There are fewer massive Cepheids than smaller mass Cepheids because of the shape of the initial mass function, hence the smaller number of massive Cepheids contribute little to the resulting fraction of Cepheids with positive or negative period change.  

This mass loss is consistent with measurements of the Cepheid mass discrepancy by \cite{Keller2008}, if we were also include some convective core overshooting.  For instance, measurements of the Large Magellanic Cloud eclipsing binary OGLE-LMC-CEP0226 \citep{Piet2010} suggest that the Cepheid component has convective core overshooting with $\alpha_c = 0.2$ \citep{Cassisi2011, Neilson2012b, Prada2012}.  

While mass-loss rates of $10^{-7}~M_\odot$~yr$^{-1}$ are consistent with observations, it is unclear how a Cepheid loses mass at such high rates.  Radiative-driven wind theory predicts rates $<10^{-10}~M_\odot$~yr$^{-1}$ \citep{Neilson2008} yet Cepheids are too hot for dust-driven winds to form.   \cite{Neilson2008,Neilson2009a} suggested an analytic model for pulsation-driven mass loss in Cepheids but the predicted mass-loss rates may still be too small to fit more recent observations, in particular $\delta$ Cephei \citep{Marengo2010, Matthews2012}.  A new description for Cepheid mass loss is required to better agree with mass-loss rates inferred from observations as well as to better understand measured rates of period change.

We demonstrate that mass loss is important for Cepheid evolution and thus might also affect evolution along the asymptotic giant branch.  While our analysis is primarily applicable to intermediate mass stars, significant mass loss in massive stars, $M > 8~M_\odot$, might lead to lower mass supernova progenitors \citep{Eldridge2008, Maund2011, Georgy2012}, and could change the blue-to-red supergiant ratio that is not understood from theoretical models \citep{Maeder2011, Langer2012}.  Along with impacting stellar evolution, Cepheid mass loss may also affect  the Cepheid period-luminosity relation. \cite{Neilson2010} suggested that infrared excess caused by mass loss can change the relation by 1-2\%.  Understanding the physical mechanism for mass loss in Cepheids is important for constraining the distance scale and for understanding stellar evolution of intermediate and massive stars.

\acknowledgements
HRN is grateful for funding provided by the Alexander von Humboldt Foundation and the National Science Foundation, grant AST-0807664. EG and SE are grateful for the support from NASA Grants HST-GO-11726.01, HST-GO-12302.01, NNX08AX37G and NASA/JPL Grant NO. 40968. HRN would like to thank Nancy R. Evans, Massimo Marengo and Douglas Welch for helpful conversations regarding mass-loss observations and David Turner for providing period change data. We thank the referee for his/her helpful comments.

%\bibliographystyle{apj}
%\bibliography{pdot}
\end{document}